\documentclass[twoside]{article}
\usepackage{qic,epsfig}
\usepackage{amsmath}
\usepackage{amsfonts}
\usepackage{amssymb}
\newcommand{\ket}[1]{|#1\rangle}
\newcommand{\bra}[1]{\langle #1|}
\newcommand{\Id}{\hat{1}}
\newcommand{\Tr}{\text{Tr}}

\begin{document}
\setlength{\textheight}{8.0truein}
\runninghead{Relative state measures of correlations in bipartite
quantum systems}{Pierre Rudolfsson, Erik Sj\"oqvist}
\normalsize\textlineskip
\thispagestyle{empty}
\setcounter{page}{1}
\copyrightheading{0}{0}{2003}{000--000}
\vspace*{0.88truein}
\alphfootnote
\fpage{1}
\centerline{\bf RELATIVE STATE MEASURES OF CORRELATIONS}
\vspace*{0.035truein}
\centerline{\bf IN BIPARTITE QUANTUM SYSTEMS}
\vspace*{0.37truein}
\centerline{\footnotesize PIERRE RUDOLFSSON}
\vspace*{0.015truein}
\centerline{\footnotesize\it Department of Quantum Chemistry, Uppsala University, Box 518,}
\baselineskip=10pt
\centerline{\footnotesize\it SE-751 20 Uppsala, Sweden}
\vspace*{10pt}
\centerline{\footnotesize ERIK SJ\"OQVIST}
\vspace*{0.015truein}
\centerline{\footnotesize\it Department of Quantum Chemistry, Uppsala University, Box 518,}
\baselineskip=10pt
\centerline{\footnotesize\it SE-751 20 Uppsala, Sweden}
\vspace*{0.015truein}
\centerline{\footnotesize\it Centre for Quantum Technologies, National University of Singapore,}
\baselineskip=10pt
\centerline{\footnotesize\it 3 Science Drive 2, 117543 Singapore, Singapore}
\vspace*{10pt}
\vspace*{0.225truein}
\publisher{(received date)}{(revised date)}
\vspace*{0.21truein}
\abstracts{Everett's concept of relative state can be viewed as a map that contains information
about correlations between measurement outcomes on two quantum systems. We demonstrate
how geometric properties of the relative state map can be used to develop operationally
well-defined measures of the total correlation in bipartite quantum systems of
arbitrary state space dimension. These measures are invariant under local unitary
transformations and non-increasing under local operations. We show that some known
correlation measures have a natural interpretation in terms of relative states.}{}{}
\vspace*{10pt}
\vspace*{3pt}
\communicate{to be filled by the Editorial}
\vspace*{1pt}\textlineskip
\vspace*{-0.5pt}
\noindent
\section{Introduction}
Ever since the formulation of the EPR argument \cite{einstein35}, the predicted
correlations between outcomes of localized quantum tests have been considered
a distinctive and important feature of quantum mechanics, with bearings on both
interpretative issues \cite{schrodinger35,bell66,pawlowski08} and possible applications
\cite{ekert91,bennett93}. There are several open problems related to correlations
within the quantum mechanical framework. The most important one is probably
the qualitative question whether a given $n$-partite state is separable or
entangled, i.e., if the correlations between the subsystems can be prepared
 by local operations and classical communication, or if global unitary evolution
(or a source of shared entanglement) is required. Related to this question is its
quantitative counterpart: how much classical correlation and entanglement does
a quantum state contain? This question has resulted in proposed measures of
correlation and entanglement, which can be divided roughly into two categories:
one that focuses on the violation of Bell-CHSH type inequalities \cite{bell65,clauser69},
and another that quantifies the ability of states to serve as
a resource in some communication task, e.g., entanglement of formation
\cite{bennett96a} and distillable entanglement \cite{bennett96b}.

One of the contexts in which quantum correlations play a significant role is
quantum measurement theory. The measurement process may be analyzed in terms
of the correlations in a closed composite system consisting of a system of
interest $S$ and an apparatus $A$. If we denote the basis states of $S$ and
$A$ by $\ket{s_i}$ and $\ket{a_i}$, respectively, and if the former
initially is in the superposition $\alpha\ket{s_0} + \beta\ket{s_1}$, then
the measurement can be described in terms of a unitary evolution resulting
in the transformation $\left( \alpha\ket{s_0} + \beta\ket{s_1} \right) \otimes
\ket{a_0} \mapsto \alpha \ket{s_0} \otimes \ket{a_0} + \beta \ket{s_1} \otimes
\ket{a_1}$. The entangled state of $S$ and $A$ corresponds to a superposition
of the possible apparatus states, which seems to be in contradiction with the definite
outcome presented by the apparatus. Everett's ``relative state formulation of
quantum mechanics'' \cite{everett57} provides a framework to deal with the
$S+A$ correlation and circumvent the ``measurement problem''. It introduces a
natural ``if - then'' perspective, equivalent to that of conditional probabilities:
{\it if} we observe the outcome $a_0$ $(a_1)$ {\it then} the state of $S$ is $s_0$
$(s_1)$, and, according to Everett, that is all there is to know. Mathematically,
this may be understood as a map from the space of apparatus states to that of the
system of interest. In this framework, the entangled state is a representation of
the relation between the possible outcomes in one measurement to those of another.
This makes the relative state formalism and the notion of conditional states potentially
useful to study correlations encoded in quantum states, as shown in e.g. the context of
entanglement  \cite{arens00} and steerability \cite{wiseman07}.

The purpose of this paper is to develop operationally well-defined correlation
measures for arbitrary bipartite states by using certain geometric properties of
the corresponding relative state map. For pure states, these measures coincide
with known entanglement measures such as concurrence hierarchies \cite{fan03} and
$I$ concurrence \cite{rungta01}. We extend these pure state measures to arbitrary
mixed bipartite systems for which we obtain measures that are invariant under local
unitary transformations as well as non-increasing under local operations. On the
other hand, these measures may increase under local operations and classical
communication (LOCC), a feature that reflects the fact that they quantify the total
correlation in mixed quantum states.

This paper is organized as follows. In the next section, we introduce the concept
of relative states in Hilbert space and operator formalisms. While the former
framework is restricted to pure bipartite states, the latter allows for an extension
of the relative state description to arbitrary mixtures of bipartite states. In section
\ref{sec:corr}, we demonstrate how to exploit the relative state concept to quantify
correlations in bipartite quantum systems of arbitrary Hilbert space dimension. The
correlation measures are illustrated in section \ref{sec:application}. The
paper ends with the conclusions.

\section{Relative states}
\subsection{Hilbert space formalism}
Let a bipartite system $S$ consisting of subsystems $A$ and $B$ be in a pure state
$\ket{\psi} \in \mathcal{H}_A\otimes\mathcal{H}_B$. Let $\dim\mathcal{H}_A =
\dim \mathcal{H}_B = d$ and $\ket{\psi} = \sum_{ij}^d \alpha_{ij}\ket{ij}$,
where $\{\ket{ij}\}$ is a product basis of the joint state space\footnote{If
$d_A = \dim \mathcal{H}_A < d_B = \dim \mathcal{H}_B$ there exists a Schmidt
decomposition with a maximum of $d_A$ components, hence the state is effectively
a $d_A\otimes d_A$.}. Following Refs. \cite{arens00,kurucz01,kurucz03}, $\ket{\psi}$ defines the
relative state map $L_\psi: \mathcal{H}_A \mapsto \mathcal{H}_B$. The relative state
operator $L_\psi$, taking a state $\ket{\varphi} \in\mathcal{H}_A$ to a state
$\ket{\phi} \in\mathcal{H}_B$ according to
\begin{eqnarray}
L_\psi \ket{\varphi} = \bra{\varphi} \psi \rangle = \ket{\phi} ,
\end{eqnarray}
may be viewed as a partial scalar product, i.e., $\bra{\varphi}\psi \rangle\equiv
(\bra{\varphi} \otimes \Id_B)\ket{\psi}$. The relative state operator can be expressed as
$L_\psi = \hat{\alpha} T$, with $\hat{\alpha} = \sum_{ij} \alpha_{ji} \ket{i} \bra{j}$
and $T$ denotes complex conjugation in the $\{ \ket{k} \}$ basis. The map
$L_\psi$ is anti-linear, i.e.,
\begin{eqnarray}
L_\psi (a\ket{\varphi_1} + b\ket{\varphi_2}) = a^{\ast} L_\psi \ket{\varphi_1} +
b^{\ast} L_\psi \ket{\varphi_2},
\end{eqnarray}
and becomes anti-unitary in the case of a maximally entangled $\ket{\psi}$. Furthermore,
$L_\psi^\dag : \mathcal{H}_B \mapsto \mathcal{H}_A$ such that $L_\psi L_\psi^{\dagger} =
\Tr_A \ket{\psi} \bra{\psi} = \rho_B$ and $L_\psi^{\dagger} L_\psi = \Tr_B
\ket{\psi} \bra{\psi} = \rho_A$. The state $\ket{\phi}$ is subnormalized
$\langle \phi \ket{\phi} = \bra{\varphi} \rho_A \ket{\varphi} \leq 1$.

In the following, we refer to an argument $\ket{\varphi}$ of the relative state map
as a \emph{hypo-state}, which can be understood as an actual post-measurement state of one
of the subsystems. The conditional state $\ket{\phi}$ we call \emph{re-state}, short
for a state relative to a hypo-state $\ket{\varphi}$.

The relative state map is a convenient way to express the fact that if Alice and Bob share
the above pure bipartite state $\ket{\psi}$ and Alice chooses to measure an observable $Q$
with eigenstates $\ket{\varphi_k}$, then she can, when an outcome $k$ is obtained, predict the
result of a specific projective measurement at Bob's site. If the shared state is entangled
with $d$ non-zero Schmidt coefficients, there will be a one-to-one correspondence between
states of $A$ and $B$, and each such pair is understood as a \emph{relative state}.

\subsection{Operator formalism}
A more general approach to relative states can be developed in terms of linear
maps of operators acting on Hilbert spaces $\mathcal{H}_A$ and $\mathcal{H}_B$.
This framework allows for mixed hypo-states that may arise in non-projective
measurements on one of the parties of bipartite states.

Denote by $\mathcal{B}(\mathcal{H})$ the space of Hermitian operators on $\mathcal{H}$. Let
$\mathcal{S}(\mathcal{H})$ be the space of semi-positive Hermitean operators with unit trace,
and $\mathcal{S}'(\mathcal{H})$ the non-negative cone of subnormalized density operators.
A bipartite state $\varrho\in\mathcal{S} (\mathcal{H}_A\otimes\mathcal{H}_B)$ defines a
map $\mathfrak{L}_{\varrho}: \mathcal{B}(\mathcal{H}_A) \mapsto \mathcal{B}(\mathcal{H}_B)$.
If $Y \in \mathcal{B}(\mathcal{H}_A)$, then the map is given by
\begin{eqnarray}
\mathfrak{L}_{\varrho}(Y) =
\Tr_A\left[(Y\otimes\Id_B)\varrho\right] \in \mathcal{B} (\mathcal{H}_B) .
\label{eq:mixmap}
\end{eqnarray}
For $Y$ being states $\tau$ of system $A$ (i.e., $\tau \in \mathcal{S} (\mathcal{H}_A)$),
then $\mathfrak{L}_{\varrho}: \mathcal{S}(\mathcal{H}_A) \mapsto \mathcal{S}'(\mathcal{H}_B)$
is the relative state map that takes hypo-states $\tau \geq 0$ on $\mathcal{H}_A$ to
(subnormalized) re-states $\pi \geq 0$ on $\mathcal{H}_B$. The map $\mathfrak{L}_{\varrho}$
is linear in the space of density operators, i.e., if $a,a'$ are real numbers and
$\tau,\tau' \geq 0$, then
\begin{eqnarray}
\mathfrak{L}_{\varrho}(a\tau + a'\tau') = a\mathfrak{L}_{\varrho}(\tau) +
a'\mathfrak{L}_{\varrho}(\tau') .
\end{eqnarray}
The norm of the re-state is the probability of finding the hypo-state in the global
state.

A mixed hypo-state $\tau$ can be understood as the post-measurement state resulting from
a (non-unique) set of projections obtained with certain probabilities. Alternatively, one
may interpret the relative state map $\mathfrak{L}_{\varrho}: \mathcal{S}(\mathcal{H}_A)
\mapsto \mathcal{S}'(\mathcal{H}_B)$ in terms of an outcome $E = V \sqrt{\tau}$
($\tau \geq 0$ and $V$ unitary) of a local generalized measurements on the $A$ system,
resulting in the post-measurement state $\pi = \Tr_A \left( E \otimes \Id_B \varrho
E^{\dagger} \otimes \Id_B \right) = \Tr_A \left( \tau \otimes \Id_B \varrho \right)$
of the $B$ system.

We can represent density operators and observables as elements of a real vector space
$\mathcal{V}$ and the relative state map can be expressed as a linear map of vectors.
The corresponding vector elements can be interpreted as the expectation values of
measured observables. Let $\{ K_k^A \}_{k=1}^{d_A^2} , \{K_l^B\}_{l=1}^{d_B^2}$ be
bases of Hermitian operators on $\mathcal{H}_A$ and $\mathcal{H}_B$, satisfying the
orthonormality conditions
\begin{eqnarray}
\Tr \left( K_k^A K_{k'}^A \right) = \delta_{kk'}, \ \Tr \left( K_l^B K_{l'}^B \right) =
\delta_{ll'}.
\label{eq:HSbasis}
\end{eqnarray}
A bipartite state can be expressed as
\begin{eqnarray}
\varrho = \sum_{kl}M_{kl}K^A_k\otimes K_l^B,
\end{eqnarray}
where
\begin{eqnarray}
M_{kl} = \Tr \left[ K^A_k\otimes K_l^B\varrho \right].
\label{eq:Mdef}
\end{eqnarray}
The matrix $M$ is a representation of $\varrho$ with respect to the chosen basis.
Local states $\tau, \pi$ are represented by real-valued vectors ${\bf a} \in \mathcal{V}
(\mathcal{H}_A)$ and ${\bf b} \in \mathcal{V} (\mathcal{H}_B)$ with elements
\begin{eqnarray}
a_k = \Tr \left[ K^A_k \tau \right] , \ b_l =  \Tr \left[ K^B_l \pi \right] .
\label{eq:abdef}
\end{eqnarray}
We can express the map in Eq. (\ref{eq:mixmap}) for $Y = \tau$ as
\begin{eqnarray}
\tau = \sum_j a_j K^A_j \mapsto \pi & = &
\mathfrak{L}_{\varrho} \left( \sum_j a_j K^A_j \right)
\nonumber \\
& = & \sum_{jkl} a_j M_{kl} \Tr \left[ K^A_j K^A_k \right] K^B_l =
\sum_{kl} a_k M_{kl} K_l^B .
\end{eqnarray}
Hence, $b_l = \sum_k a_k M_{kl}$ or equivalently ${\bf b} = M^{\textrm{T}}{\bf a}$.
The relative state map is represented by
\begin{eqnarray}
M^{\textrm{T}} : \mathcal{V}(\mathcal{H}_A) \mapsto \mathcal{V}(\mathcal{H}_B)
\label{eq:HSmap1}
\end{eqnarray}
and conversely
\begin{eqnarray}
M : \mathcal{V}(\mathcal{H}_B) \mapsto \mathcal{V} (\mathcal{H}_A) .
\label{eq:HSmap2}
\end{eqnarray}

\section{Correlations}
\label{sec:corr}
A probability distribution $P(X,Y)$ over two random variables  $X,Y$ taking values
$x_i,y_j$ is correlated if $ P(X,Y)\neq P(X)P(Y)$, where $P(X)$ and $P(Y)$ are the
marginal distributions of $P(X,Y)$. The above condition can also be stated in terms
of conditional probabilities: if there exist a pair $(i,j\neq i)$ such that
\begin{eqnarray}
P(X|Y = y_i) \neq P(X|Y=y_j) ,
\label{eq:corrcond}
\end{eqnarray}
where $P(X|Y=y_i)$ denotes the probability distribution over $X$ given the outcome
$Y = y_i$, then $P(X,Y)$ is correlated. The condition Eq. (\ref{eq:corrcond}) states
that a probability distribution is correlated if information about an outcome of $Y$
alters the prediction about the outcome of $X$ (and vice versa). Thus, a way to
characterize the correlation in a probability distribution is to compare the set
of conditional probabilities given an exhaustive set of mutually exclusive conditionals
$\{y_i\}$, since the set of conditional probabilities contains information about
how the random variables are correlated, i.e., which outcomes $x_i$ are
correlated with which outcomes $y_j$.

A quantum state $\varrho\in \mathcal{S}(\mathcal{H}_A\otimes\mathcal{H}_B)$ is correlated if
$\varrho\neq \rho_A\otimes\rho_B$, where $\rho_A$ and $\rho_B$ are the reduced
states of the $A$ and $B$ subsystems, respectively. The relative state formalism allows us
to employ Eq. (\ref{eq:corrcond}) in a quantum context: a bipartite state $\varrho$ is
correlated if there exists a pair of Hermitean operators $Y' \neq Y$ such that
\begin{eqnarray}
\Tr_A (Y' \otimes \Id_B \varrho) \neq \lambda \Tr_A (Y \otimes \Id_B \varrho)
\end{eqnarray}
for any real number $\lambda$. Here, the conditional probabilities are replaced by relative
states. The basic idea of the following analysis is to measure correlations in terms of how
much the re-states differ for different choices of hypo-states. Specifically, the aim is to
quantify correlations in terms of the difference of the conditional predictions contained
in the re-states. In this way, we demonstrate how the geometrical properties of the relative
state map can be used to develop correlation measures in arbitrary bipartite quantum systems.

\subsection{Pure state correlation measures}

\begin{figure} [htbp]
\centerline{\epsfig{file=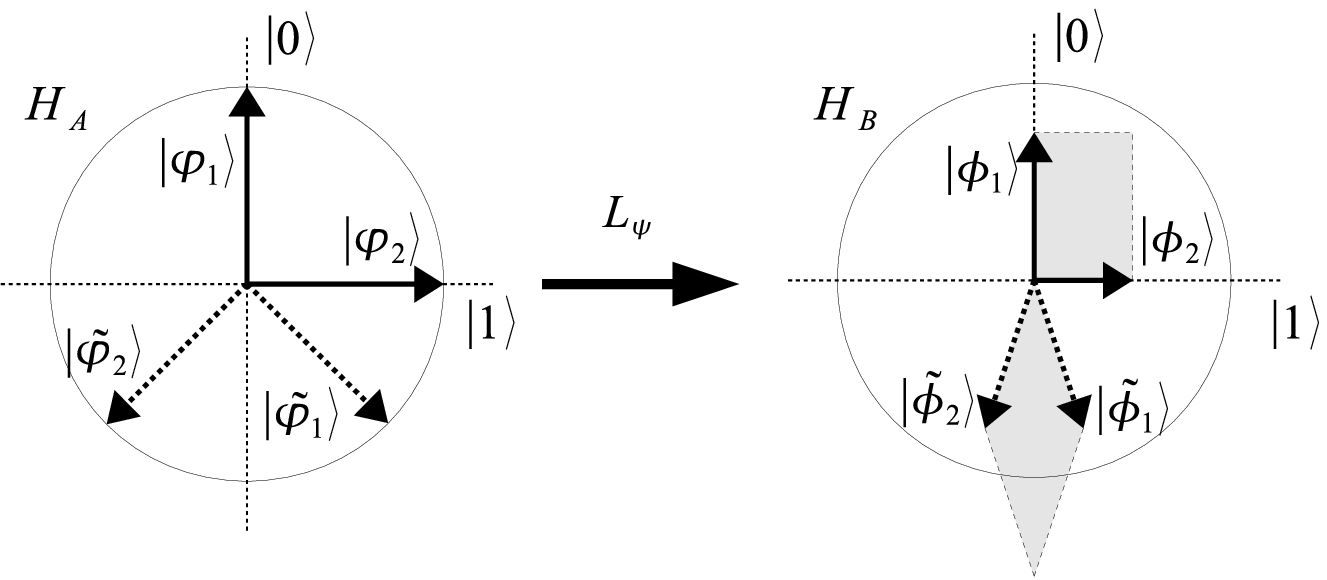, width= 10 cm}}
\vspace*{13pt}
\fcaption{\label{fig:geom2x2} An illustration of the correlation measure for a $2\otimes 2$
system in pure state $\ket{\psi} = c_0\ket{00} + c_1\ket{11}$. The picture shows the real planes
in $\mathcal{H}_A$ and $\mathcal{H}_B$ spanned by the local Schmidt bases. Two different choices
of orthonormal hypo-states in $\mathcal{H}_A$, $\ket{\varphi_i}$ and $\ket{\tilde{\varphi}_i}$,
maps to their respective re-states $\ket{\phi_i}$ and $\ket{\tilde{\phi}_i}$ in $\mathcal{H}_B$.
The areas spanned by the restates are shown in gray.}
\end{figure}

Let us consider the bipartite pure product state $\ket{\psi} = \ket{\psi_A}\otimes\ket{\psi_B}$ and
the corresponding relative state map $L_\psi$. For any hypo-state $\ket{\varphi}\in \mathcal{H}_A$,
the re-state is
\begin{eqnarray}
\ket{\phi} = \bra{\varphi} \psi \rangle = \bra{\varphi} \psi_A \rangle \ket{\psi_B},
\end{eqnarray}
i.e., $L_\psi$ maps the whole Hilbert space $\mathcal{H}_A$ to the same ray in $\mathcal{H}_B$,
that is, to the same state. This expresses the fact that for an uncorrelated state
$\ket{\psi_A}\otimes\ket{\psi_B}$,  a measurement outcome at site $A$ does not change the
predictions about measurements at site $B$. Now, consider instead an entangled two-qubit state
$\ket{\psi} \in \mathcal{H}_A \otimes \mathcal{H}_B$. If we choose two hypo-states
$\ket{\varphi},\ket{\varphi'}$ such that the corresponding re-states
$\ket{\phi}, \ket{\phi'}$ are non-zero, then $\ket{\varphi}\neq\ket{\varphi'}$ implies
that $\ket{\phi}\neq z\ket{\phi'}$, i.e., a measurement outcome at site $A$ does change
the predictions regarding measurements at site $B$, as illustrated in Fig. \ref{fig:geom2x2}.

By using these properties of the relative state map we may develop measures that
quantify bipartite correlations. These measures are based upon geometric properties
of the wedge product, $\wedge$ defined as follows. Let $\{ \ket{j} \}_{j=1}^d$ be an
orthonormal basis of a $d$ dimensional Hilbert space $\mathcal{H}$. Consider the
vectors $\ket{\xi_i} = \sum_{j=1}^d \eta_j^{(i)} \ket{j}$, $i=1,\ldots ,k\leq d$.
We define the $k$-product of these vectors as
\begin{eqnarray}
\xi_{1} \wedge \cdots \wedge \xi_{k} \sim \ket{\xi_{1}} \wedge \cdots \wedge \ket{\xi_{k}}
\equiv \sum_{1\leq \mu_1 < \ldots < \mu_k \leq d} \sum_{\ j_1 \ldots j_k}
\epsilon_{j_1 \ldots j_k}^{\mu_1 \ldots \mu_k}
\eta_{j_1}^{(1)} \cdots \eta_{j_k}^{(k)} \ket{\mu_1 \ldots \mu_k} ,
\end{eqnarray}
where $\epsilon_{j_1 \ldots j_k}^{\mu_1 \ldots \mu_k}$ is the Levi-Civita tensor,
defined as $\epsilon_{j_1 \ldots j_k}^{\mu_1 \ldots \mu_k} = +1 \ (-1)$ if
$j_1 \ldots j_k$ is an even (odd) permutation of $\mu_1 \ldots \mu_k$ and
zero otherwise. Note in particular that the $k$-product vanishes if $\xi_i$
are linearly dependent. $\xi_1 \wedge \cdots \wedge \xi_k$ is an element of
the exterior space $\Omega^k (\mathcal{H})$ with norm
\begin{eqnarray}
\left| \xi_{1} \wedge \cdots \wedge \xi_{k} \right|^2
\equiv \sum_{1\leq \mu_1 < \ldots < \mu_k \leq d}
\left| \sum_{\ j_1 \ldots j_k } \epsilon_{j_1 \ldots j_k}^{\mu_1 \ldots \mu_k}
\eta_{j_1}^{(1)} \cdots \eta_{j_k}^{(k)} \right|^2 .
\end{eqnarray}

\begin{figure} [htbp]
\centerline{\epsfig{file=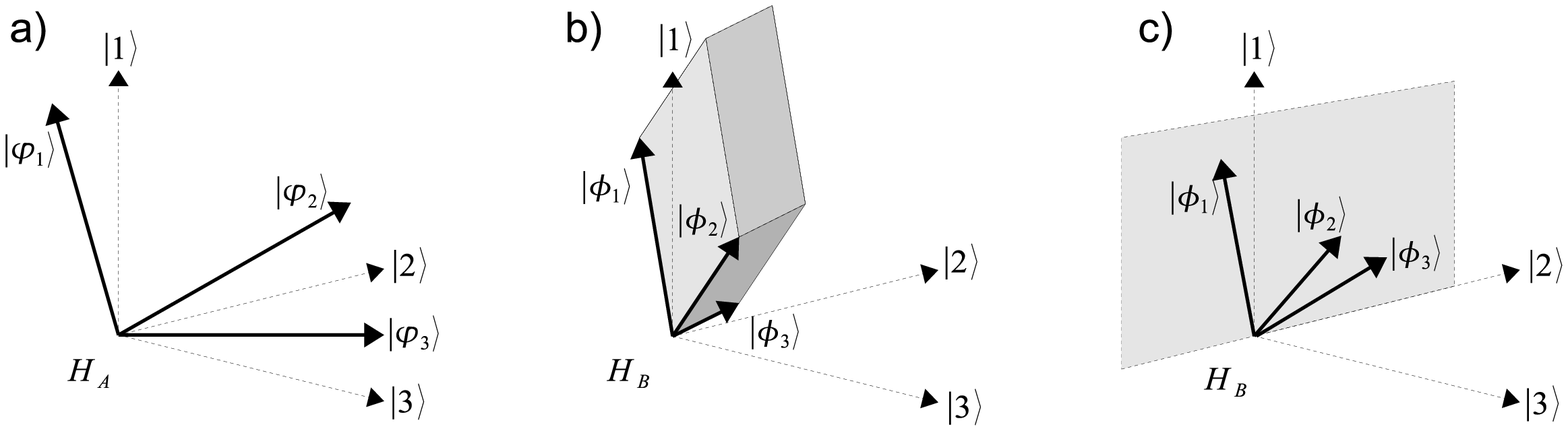, width=14 cm}}
\vspace*{13pt}
\fcaption{\label{fig:geom3x3} An illustration of the correlation measures $\lambda_{ij}, \lambda_{123}$
for a $3\times 3$ system in two pure states with different Schmidt-number, $\ket{\psi} = \sum_{k=1}^3
c_k\ket{kk}$ and $\ket{\widetilde{\psi}} = \sum_{k=1}^2\widetilde{c}_k\ket{kk}$. In a) a choice of
hypostates $\ket{\varphi_i}\in\mathcal{H}_A$ is depicted, and b) shows the re-states
$L_\psi \ket{\varphi_i} = \ket{\phi_i}\in\mathcal{H}_B$, which span the volume $\lambda_3 =
|\phi_1\wedge\phi_2\wedge\phi_3|$ taken as the measure of three-level correlations. The areas
of the faces of the rhomboid are given by $\lambda_{12} = |\phi_1\wedge\phi_2|$, $\lambda_{13} =
|\phi_1\wedge\phi_3|$ and $\lambda_{23} = |\phi_2\wedge\phi_3|$, and they are measures of the
two-level correlations bewteen the respective two-dimensional subspaces. In c) the re-states
$L_{\widetilde{\psi}}\ket{\varphi_1} = \ket{\widetilde{\phi}_i}$ are shown, which lies in the
subspace (shown in gray) spanned by the Schmidt vectors $\ket{1},\ket{2}$. Consequently, for
$\ket{\widetilde{\psi}}$, the three-level correlations are $\lambda_{123} = |\widetilde{\phi}_1
\wedge\widetilde{\phi}_2\wedge\widetilde{\phi}_3| = 0$, whereas $\lambda_{ij} \neq 0$, i.e.,
$\ket{\widetilde{\psi}}$ only contains two-level correlations.  }
\end{figure}

For a set of vectors $\left\{ \mathbf{v}_i \right\}$ in a real three dimensional
vector space $\mathcal{V}^3$, the two-fold wedge product $\mathbf{v}_{ij} =
\mathbf{v}_i\wedge \mathbf{v}_j\in\Omega^2(\mathcal{V}^3)$ can be identified with
the directed surface element spanned by the two vectors, with area $|\mathbf{v}_{ij}|$.
Correspondingly, the three-fold wedge product $\mathbf{v}_{ijk} = \mathbf{v}_i \wedge
\mathbf{v}_j \wedge \mathbf{v}_k \in \Omega^3(\mathcal{V}^3)$ represents a directed
volume element, with volume $|\mathbf{v}_{ijk}|$ (see Fig. \ref{fig:geom3x3} for an
illustration). This geometric intuition carries over to complex higher dimensional
spaces; the $k$-fold wedge product $\xi_{i_1...i_k} = \xi_{i_1}\wedge\cdots\wedge\xi_{i_k}$
can be seen as the oriented $k$-dimensional rhomboid spanned by the vectors,
with $k-$volume $|\xi_{i_1...i_k}|$.

Given a general bipartite system prepared in the pure state $\ket{\psi} \in
\mathcal{H}_A\otimes\mathcal{H}_B$, where we assume that $\dim\mathcal{H}_A =
\dim\mathcal{H}_B = d$, a set of hypo-states $\{\ket{\varphi_i}\}_{i=1}^d$,
$\ket{\varphi_i}\in\mathcal{H}_A$, is chosen such that $\mathrm{Sp} \{\ket{\varphi_i}\}
\cong \mathcal{H}_A$. We obtain a set of re-states $\{\ket{\phi_i}\}_{i=1}^d$,
$\ket{\phi_i}\in\mathcal{H}_B$, via $ \ket{\phi_i} = L_\psi \ket{\varphi_i} =
\bra{\varphi_i} \psi \rangle$. Our basic measure of correlation with respect to any
$k$-tuple of hypo-states $\{\ket{\varphi_{i_1}}, \ldots ,\ket{\varphi_{i_k}}\} \subseteq
\{\ket{\varphi_i}\}_{i=1}^d$ is given by
\begin{eqnarray}
\lambda_{i_1 \ldots i_k} = \frac{|\phi_{i_1} \wedge \cdots \wedge \phi_{i_k}|}
{| \varphi_{i_1}\wedge \cdots \wedge \varphi_{i_k}|} .
\label{eq:entmeas}
\end{eqnarray}
We may interpret $\lambda_{i_1 \ldots i_k}$ as follows. Each $k$-tuple
$\{\ket{\varphi_{i_1}}, \ldots ,\ket{\varphi_{i_k}}\} \subseteq \{\ket{\varphi_i}\}_{i=1}^d$
of hypo-states is a basis of a $k$-dimensional subspace $\mathcal{H}_A^{i_1 \ldots i_k}
\subseteq \mathcal{H}_A$, and we will call $\lambda_{i_1 \ldots i_k}$ a measure of the $k$-level
correlation between that subspace and subsystem $B$.

To make this notion clearer, consider a $3\otimes 3$ system in the state $\ket{\psi} = \sum_{i=1}^3\sqrt{p_i}\ket{ii}$, and a choice of hypo-states as $\ket{\varphi_i} = \ket{i}$, with the corresponding re-states given by $\ket{\phi_i} = L_\psi\ket{\varphi_i} = \sqrt{p_i}\ket{i}$. By Eq. (\ref{eq:entmeas}), we have three quantities for the two-level correlations $\lambda_{ij} = |\phi_i\wedge\phi_j| = \sqrt{p_ip_j},\ i<j$, and one for the three-level correlation $\lambda_{123} = |\phi_1\wedge\phi_2\wedge\phi_3| = \sqrt{p_1p_2p_3}$, see Fig. \ref{fig:geom3x3}b where a similar example with a different choice of hypo-states is illustrated. The quantity $\lambda_{13} = \sqrt{p_1p_3}$ quantifies the difference between the restates $\ket{\phi_1},\ket{\phi_3}$, and hence corresponds to how much our predictions about measurements on system $B$ differs with the two post-measurement states $\ket{\varphi_1},\ket{\varphi_3}$ of $A$, i.e., when the outcome corresponding to $\ket{\phi_2}$ is discarded. Equivalently, $\lambda_{13}$ measures the effective $2\otimes 2$ entanglement in the state $\ket{\psi'} = (\ket{1}\bra{1} +\ket{3}\bra{3} )\otimes(\ket{1}\bra{1} + \ket{3}\bra{3} )\ket{\psi}$ resulting from a projection onto the subspace $\mathcal{H}^{13}_{A}\otimes\mathcal{H}^{13}_{B}$. The three-level quantity $\lambda_{123}$ measures the volume spanned by the re-states, i.e., how much the predictions differ when all three post-measurement states $\ket{\varphi_i}$ are taken into account. On the other hand, if $p_3 = 0$, then $\lambda_{13} = \lambda_{23} = \lambda_{123} = 0$, where $\lambda_{13} = \lambda_{23} = 0$ reflects that $\ket{\psi'} = \sqrt{p_1}\ket{11}$ is a product state (the subspace $\mathcal{H}^{13}_{A}$ is not correlated with $B$), and $\lambda_{123} = 0$ means that there exist no correlations that is not two-level. As is shown in Fig. \ref{fig:geom3x3}c, the linear dependence of the re-states tells us that the information in, e.g., $\ket{\phi_3}$, is already present in $\ket{\phi_1},\ket{\phi_2}$. Note that the denominator of Eq. (\ref{eq:entmeas}) can be viewed as a normalization factor quantifying how much the hypo-states differ in the first place.

The quantity $\lambda_{i_1 \ldots i_k}$ is independent of the choice of hypo-states as long
as $\mathrm{Sp}\left\{ \ket{\varphi_{i_1}}, \ldots ,\ket{\varphi_{i_k}} \right\}\cong
\mathcal{H}_{i_1 \ldots i_k}$. To see this, let $\{\ket{\varphi_{i_l}}\}_{l=1}^k$ form
an orthonormal basis of $\mathcal{H}_{i_1 \ldots i_k}$ and define another set
$\{\ket{\varphi'_{i_l}}\}_{l=1}^k$ of arbitrary basis vectors via
\begin{eqnarray}
\ket{\varphi'_{i_l}} = \sum_m c_{lm}\ket{\varphi_{i_m}},
\label{newhypos}
\end{eqnarray}
where $c_{lm}$ are elements of a complex-valued invertible $k \times k$ matrix.
Define the corresponding set of re-states $\{ \ket{\phi'_{i_l}} \}_{l=1}^k$ as
$\ket{\phi'_{i_l}} = L_\psi\ket{\varphi'_{i_l}}$. By the anti-linearity of the
relative state map, we obtain
\begin{eqnarray}
\ket{\phi'_{i_l}} = L_\psi\left(\sum_m  c_{lm} \ket{\varphi_{i_m}}\right) =
\sum_m c_{lm}^{\ast} \ket{\phi_{i_m}}.
\end{eqnarray}
Explicit evaluation of the wedge product of the non-orthogonal basis elements yields
\begin{eqnarray}
\varphi'_{i_1} \wedge \cdots \wedge \varphi'_{i_k}
& = & \left(\sum_{j_1 \ldots j_k} \epsilon_{j_1 \ldots j_k} c_{1 j_1} \cdots
c_{k j_k}\right) \varphi_{i_1} \wedge \cdots \wedge \varphi_{i_k} ,
\label{eq:levicivita1}
\end{eqnarray}
where $\epsilon_{\nu_1 \ldots \nu_d}$ denotes the Levi-Civita tensor. By performing
the corresponding expansion of the set of primed re-states $\phi_{\mu}'$, we obtain
\begin{eqnarray}
\phi'_{i_1} \wedge \cdots \wedge \phi'_{i_k}
& = & \left(\sum_{j_1 \ldots j_k} \epsilon_{j_1\ldots j_k} c_{1 j_1}^{\ast} \cdots
c_{k j_k}^{\ast}\right) \phi_{i_1} \wedge \cdots \wedge \phi_{i_k} ,
\label{eq:levicivita2}
\end{eqnarray}
which is essentially the same expression as in Eq. (\ref{eq:levicivita1}) up to a complex
conjugation of the coefficient $c_{lm}$. Thus, we conclude that
\begin{eqnarray}
|\phi_{i_1} \wedge \cdots \wedge \phi_{i_k}| = \frac{|\phi'_{i_1} \wedge \cdots
\wedge \phi'_{i_k}|}{|\varphi'_{i_1} \wedge \cdots \wedge \varphi'_{i_k}|}.
\end{eqnarray}
To simplify the notation, we henceforth assume that a set of hypo-states form
an orthogonal basis and thus omit the denominator in Eq. (\ref{eq:entmeas}).

The correlation quantities $\lambda_{i_1 \ldots i_k}$ have the following properties.
If the re-states $(\phi_{i_1}, \ldots ,\phi_{i_k})$ are linearly dependent, then
$\lambda_{i_1 \ldots i_k} = 0$, which reflects that one can find two hypo-states in
$\mathcal{H}_{i_1 \ldots i_k}$ that maps to the same ray in $\mathcal{H}_B$. Furthermore,
$\lambda_{i_1 \ldots i_k}$ vanishes if $\ket{\psi}$ lacks support in some part of the
subspace spanned by the hypo-states (one or several re-states will have zero norm).
On the other hand, $\max \lambda_{i_1 \ldots i_k} = (1/\sqrt{k})^k$ and this value is
saturated if $\ket{\psi}$ is maximally entangled on $\mathcal{H}_{i_1 \ldots i_k}$ and
$\ket{\varphi_i}, \ i \in (i_1 \ldots i_k)$ span this subspace.

The quantities $\lambda_{i_1 \ldots i_k}$ are in general not invariant under local unitaries.
To see this, consider the local unitary transformation $\ket{\psi} \mapsto \ket{\psi'} =
U_A\otimes\Id_B\ket{\psi}$, which implies that $\ket{\phi_i} \mapsto \ket{\phi'_i} =
\bra{\varphi_i}U_A\otimes\Id_B\ket{\psi}$. In other words, the transformed re-states
would correspond to a set of hypo-states $\ket{\varphi'_i} = U_A^{\dagger}
\ket{\varphi_i}$, defining a different subspace decomposition of $\mathcal{H}_A$ leading
to that $\lambda_{i_1 \ldots i_k}$ may change. (The exception is $\lambda_d =
\phi_1 \wedge \cdots \wedge \phi_d$ that contains all re-states.) However, we have seen
that the different choices of orthonormal bases of hypo-states are equivalent with local
unitary transformations (on subsystem A) of the global state, and hence the question of
invariance under change of hypo-states are equivalent to that of invariance under local
unitary transformations. We now define
\begin{eqnarray}
\Lambda^2_k = d^k \binom{d}{k}^{-1}
\sum_{i_1< \ldots <i_k}^d \lambda_{i_1 \ldots i_k}^2 ,
\end{eqnarray}
where the sum is over all unique $k$-tuples of re-states and the normalization
factor on the right-hand side is chosen so that $\Lambda_k = 1$ for all $k$
if the global state is maximally entangled.

{\it Theorem.} For a $d\times d$-dimensional bipartite system, the members of the
set $\{ \Lambda_k \}_{k=1}^d$ are invariant under local unitary transformations.

{\it Proof.} Let $\ket{\psi} = \sum_{i=1}^d\sqrt{p_i}\ket{ii}$ be the bipartite state
on Schmidt form. We can make a choice of hypo-states such
that $\ket{\varphi_i}  =  \ket{i}$, with the corresponding re-states $\ket{\phi_i} =
L_\psi\ket{\varphi_i} = \sqrt{p_i}\ket{i}$. First, we consider a local unitary on subsystem
$B$, i.e.,
\begin{eqnarray}
\ket{\psi} \mapsto \ket{\widetilde{\psi}} = \Id_A\otimes U_B\ket{\psi} =
\sum_{i=1}^d\sqrt{p_i}\ket{i}\otimes U_B\ket{i},
\end{eqnarray}
from which we se that the re-states transform according to $\ket{\phi_i} \mapsto
\ket{ \widetilde{\phi}_i } = U_B\ket{\phi_i}$. The corresponding transformation of
the $k$-vectors then reads
\begin{eqnarray}
\phi_{i_1} \wedge\cdots\wedge\phi_{i_i} \mapsto \widetilde{\phi}_{i_1} \wedge \cdots \wedge
\widetilde{\phi}_{i_k} = U_B^{\otimes k} \left( \phi_{i_1} \wedge \cdots \wedge \phi_{i_k} \right),
\end{eqnarray}
which means that the unitary $U_B$ on $\mathcal{H}_B$ induces a unitary $U_B^{\otimes k}$ on
the exterior space $\Omega^k(\mathcal{H}_B)$. Clearly, this cannot change the norm of the
$k$-vector, since
\begin{eqnarray}
\widetilde{\lambda}_{i_1 \ldots i_k}^2 = \left(\phi_{i_1} \wedge\cdots\wedge\phi_{i_k}\right)^{\dagger}
\left(U_B^{\otimes k}\right)^{\dagger} U_B^{\otimes k}\left(\phi_{i_1} \wedge\cdots\wedge\phi_{i_k}
\right) = \left| \phi_{i_1} \wedge \cdots \wedge \phi_{i_k} \right|^2 = \lambda_{i_1 \ldots i_k}^2,
\end{eqnarray}
and thus we have that $\widetilde{\Lambda}_k = \Lambda_k$ under $U_B$.

Consider now a local unitary on subsystem $A$, i.e., let $\ket{\psi}$ be defined as before but
let $\ket{\psi}\mapsto\ket{\widetilde{\psi}} = U_A\otimes\Id_B\ket{\psi}$. In this case, the
re-states transform as
\begin{eqnarray}
\ket{\phi_i}\mapsto\ket{ \widetilde{\phi}_i } = L_{\widetilde{\psi}}\ket{\varphi_i} =
\bra{\varphi_i}U_A\otimes\Id_B\ket{\psi} =
L_\psi U_A^\dag\ket{\varphi_i} = L_\psi\ket{\widetilde{\varphi_i}},
\end{eqnarray}
i.e, a local unitary on subsystem $A$ is equivalent to the inverse transformation of the hypo-states.
If we denote $(U_A^\dag)_{ij} = u_{ij}$, the transformed hypo-states are related to the original ones
according to $\ket{\widetilde{\varphi}_j} = \sum_iu_{ij}\ket{\varphi_i}$, and we have that
$\ket{\widetilde{\phi}_j} = L_\psi\ket{\widetilde{\varphi}_j}=  \sum_iu^*_{ij}L_\psi\ket{\varphi_i} =
\sum_iu^{\ast}_{ij}\ket{\phi_i} $. To show that $\widetilde{\Lambda}_k = \Lambda_k$, we first
note that the $U_A$ induces a corresponding transformation of the $k$-vectors
\begin{eqnarray}
\phi_{j_1}\wedge\cdots\wedge\phi_{j_k} \mapsto \widetilde{\phi}_{j_1} \wedge \cdots \wedge
\widetilde{\phi}_{j_k} = \left( \sum_{i_1} u_{i_1j_1}^{\ast} \phi_{i_1} \right) \wedge \cdots
\wedge \left( \sum_{i_k} u_{i_kj_k}^{\ast} \phi_{i_k} \right),
\end{eqnarray}
and summing the squared norms of the new $k$-vectors, we get
\begin{eqnarray}
\sum_{1\leq j_1 < \ldots < j_k\leq d}\left|\widetilde{\phi}_{j_1} \wedge \cdots \wedge
\widetilde{\phi}_{j_k} \right|^2 =
\nonumber &\\
\sum_{\substack{1\leq j_1< \ldots <j_k\leq d \\
1\leq\mu_1< \ldots <\mu_k\leq d \\ i_1 \ldots i_k\\
m_1 \ldots m_k}}
\epsilon_{i_1 \ldots i_k}^{\mu_1 \ldots \mu_k} \epsilon_{m_1 \ldots m_k}^{\mu_1 \ldots \mu_k} & u_{i_1j_1}^{\ast}
u_{m_1j_1} \cdots u_{i_kj_k}^{\ast} u_{m_1j_1} \left| \phi_{\mu_1} \wedge \cdots \wedge \phi_{\mu_k}
\right|^2.
\end{eqnarray}
Here, we have used that the set $\left\{ \phi_{\mu_1} \wedge \cdots \wedge \phi_{\mu_k}
\right\}_{1\leq\mu_1<...<\mu_k\leq d}$ is an orthogonal basis of $\Omega^k(\mathcal{H}_B)$,
which follows from the orthogonality of the Schmidt-basis and that $\ket{\phi_i} = \sqrt{p_i}\ket{i}$.
Now, to see that the factors labeled by $\mu_1...\mu_k$ each sum up to one as required, first note
that the rows of a unitary is an orthonormal set of vectors, i.e., we have that $\sum_j
u^*_{\mu_i j}u_{\mu_m j} = \delta_{im}$. The determinant of the identity can then be expanded
according to
\begin{eqnarray}
1 = \det \delta_{im} & = & \frac{1}{k!}\sum_{\substack{j_1...j_k \\
i_1...i_k\\
m_1...m_k}}\epsilon_{i_1...i_k} \epsilon_{m_1...m_k} u^*_{\mu_{i_1}j_1}u_{\mu_{m_1}j_1}
\cdots u^*_{\mu_{i_k}j_k}u_{\mu_{m_k}j_k}
\nonumber\\
& = & \sum_{\substack{1\leq j_1<...<j_k\leq d \\
i_1...i_k\\
m_1...m_k}} \epsilon_{i_1...i_k}^{\mu_1...\mu_k}\epsilon_{m_1...m_k}^{\mu_1...\mu_k}
u_{i_1j_1}^*u_{m_1j_1}\cdots u_{i_kj_k}^*u_{m_kj_k},
\end{eqnarray}
where we have used the definition of the Levi-Civita tensor and that we can restrict
the sums over $j_1...j_k$.
\begin{flushright}
$\Box$
\end{flushright}

The $k=1$ invariant is just normalization and does not provide any information about the
correlation between the subsystems. Therefore, we take the correlation measures to consist
of the set $\{ \Lambda_k \}_{k=2}^d$.

Note that the $k$-level invariants are not independent since $\Lambda_k \neq 0$ implies
that $\Lambda_l \neq 0$ for all $l<k$. Geometrically, this expresses the fact that a non-zero
volume must be bounded by non-zero areas. More explicitly, the lower order invariants are
related to $\Lambda_d$ as
\begin{eqnarray}
\Lambda_k \geq \binom{d}{k}^{1/2} (\Lambda_d)^{k/d},\ d>k,
\end{eqnarray}
which gives a lower bound for $k$th order invariant.

We may relate the $\Lambda_k$'s to known entanglement measures by using the Schmidt
form $\ket{\psi} = \sum_{k=1}^d \sqrt{p_k}\ket{\varphi_k}\otimes\ket{\phi_k}$, where
$\langle \varphi_k \ket{\varphi_l} = \delta_{kl}$ and $\langle \phi_k \ket{\phi_l} =
\delta_{kl}$, such that $L_\psi\ket{\varphi_k} = \sqrt{p_k}\ket{\phi_k}$. Since
the re-states are subnormalized, mutually orthogonal vectors, it follows that
\begin{eqnarray}
\Lambda_k^2 =
\sum_{i_1< \ldots <i_k} \left| \phi_{i_1} \wedge \cdots \wedge \phi_{i_k} \right|^2 =
\sum_{i_1< \ldots <i_k} p_{i_1} \cdots p_{i_k}.
\end{eqnarray}
Hence, the invariants are equivalent to the symmetric polynomials in the Schmidt
coefficients, i.e., the concurrence hierarchies proposed in Ref. \cite{fan03}. The
pure state invariant $\Lambda_2$ is recognized as the $I$ concurrence \cite{rungta01}
\begin{eqnarray}
C_I^2 = 4\sum_{i<j} p_i p_j.
\end{eqnarray}
up to a factor. For $d=2$ (qubit) systems, $\Lambda_2$ is the only non-trivial invariant
and equals half the pure state concurrence \cite{wootters98}.

The relative state approach may further be used to give the following alternative
geometric interpretation of pure state concurrence for qubit systems. Let
$\ket{\Psi} = \sqrt{p_0} \ket{00} + \sqrt{p_1} \ket{11}$ and consider
the orthonormal hypo-states $\ket{\varphi_0} = \alpha \ket{0} + \beta \ket{1}$ and
$\ket{\varphi_1} = -\beta^{\ast} \ket{0} + \alpha^{\ast} \ket{1}$ with complex-valued
$\alpha$ and $\beta$ such that $|\alpha|^2 + |\beta|^2 = 1$. The corresponding re-states
read $\ket{\phi_0} = \sqrt{p_0} \alpha^{\ast} \ket{0} + \sqrt{p_1} \beta^{\ast} \ket{1}
\sim {\bf a} = (\sqrt{p_0} \alpha^{\ast},\sqrt{p_1} \beta^{\ast})$ and $\ket{\phi_1} =
-\sqrt{p_0} \beta \ket{0} + \sqrt{p_1} \alpha \ket{1} \sim {\bf b} = (- \sqrt{p_0}\beta ,
\sqrt{p_1} \alpha)$. The area $\mathcal{A}$ spanned by ${\bf a}$ and ${\bf b}$ is
\begin{eqnarray}
\mathcal{A} = \sqrt{|{\bf a}|^2 |{\bf b}|^2 -
\left| {\bf a}^{\ast} \cdot {\bf b} \right|^2} = \sqrt{p_0 p_1} ,
\end{eqnarray}
which is half the pure state concurrence of the two-qubit state $\psi$. Thus,
concurrence is essentially the area spanned by two re-states, as is shown in
Fig. \ref{fig:geom2x2}.

\subsection{Mixed state correlation measures}
Let $\varrho\in \mathcal{S}(\mathcal{H}_A\otimes\mathcal{H}_B)$ be a bipartite state and
assume that $d=\dim\mathcal{H}_A \leq \dim \mathcal{H}_B$. Let $\{\tau_i\}_{i=1}^{d^2},\
\tau_i\in\mathcal{B}(\mathcal{H}_A)$, be a set of Hermitean operators on $\mathcal{H}_A$ such
that $\mathrm{Sp}\left\{ \tau_i \right\} \cong \mathcal{B}(\mathcal{H}_A)$ and define the
corresponding set of operators $\{\pi_i\}_{i=1}^{d^2},\ \pi_i \in \mathcal{S}'(\mathcal{H}_B)$,
as
\begin{eqnarray}
\pi_i = \mathfrak{L}_\varrho(\tau_i) = \Tr_A \left[ \tau_i\otimes\Id\varrho \right].
\end{eqnarray}
The basic correlation measures now read
\begin{eqnarray}
\upsilon_{i_1 \ldots i_k} =
\frac{|\pi_{i_1} \wedge \cdots \wedge \pi_{i_k}|}{|\tau_{i_1} \wedge \cdots \wedge
\tau_{i_k}|} .
\label{eq:entmeas2}
\end{eqnarray}
In analogy with the pure state case, these measures are independent of choice of
$\{ \tau_i \}$. In particular, if $\{\tau_i\}$ is an orthogonal set the denominator
in Eq. (\ref{eq:entmeas2})) can be omitted. Note, however, the operational
interpretation of $\{\tau_i\}$ and $\{\pi_i\}$ as states cannot be maintained for such
a choice, since the space of density operators cannot be equipped with a complete orthogonal
basis of positive operators. Nonetheless, due to the independence of the choice of
$\{ \tau_i \}$, we refer to $\{\tau_i\}$ and $\{\pi_i\}$ as hypo-states and re-states
in the following, regardless of whether all members of the sets represent valid states
or not.

To evaluate the wedge product, it is convenient to move to the Hilbert-Schmidt
representation of states and observables as real-valued vectors and matrices. Thus, we
make the substitutions $\varrho \rightarrow M$, $\{\tau_i\} \rightarrow \{ {\bf a}_i \}$, and
$\{ \pi_i \} \rightarrow \{ {\bf b}_i \}$, where ${\bf a}_i$ and ${\bf b}_i$ are
related via the linear map ${\bf a}_i \rightarrow {\bf b}_i = M^{\textrm{T}} {\bf a}_i$. Then
\begin{eqnarray}
\pi_{i_1} \wedge \cdots \wedge \pi_{i_k} \rightarrow
{\bf b}_{i_1} \wedge \cdots \wedge {\bf b}_{i_k} .
\end{eqnarray}
To illustrate this substitution, let us consider the case of a product state $\varrho =
\rho_A \otimes \rho_B$. We find $M = {\bf r}_A {\bf r}_B^{\textrm{T}}$, where
$r_{A;i} = \Tr \left[ K_i^A \rho_A \right]$ and $r_{B;i} = \Tr \left[ K_i^B \rho_B \right]$
for some local operator bases $\{ K_i^A \}$ and $\{ K_i^B \}$. Hence, for a product state,
the relative state map takes any ${\bf a} \in \mathcal{V}_A$ to a vector proportional to
${\bf r}_B$: $M^{\textrm{T}}{\bf a} = {\bf r}_B \left( {\bf r}_A^{\textrm{T}} \cdot {\bf a} \right)$,
which implies $|(M^{\textrm{T}}{\bf a}_1) \wedge \cdots \wedge (M^{\textrm{T}}{\bf a}_k)| = 0$
for any $k$-tuple $({\bf a}_1, \ldots ,{\bf a}_k) \in \mathcal{V}_A$.

We now define the correlation measures
\begin{eqnarray}
\Upsilon_k^2 = d^{2k} \binom{d^2}{k}^{-1}
\sum_{i_1 < \ldots < i_k}^{d^2} \upsilon_{i_1 \ldots i_k}^2,
\end{eqnarray}
where the normalization factor is chosen such that $\Upsilon_k = 1$
for maximally entangled states.

To demonstrate that $\Upsilon_k$ are invariant under local unitary operations,
we first need to define the corresponding transformation in the Hilbert-Schmidt
representation. Let $\varrho$ be a state and let ${\bf r}$ be the Hilbert-Schmidt
representation of the state given by $r_i = \Tr \left[ K_i\varrho \right]$. Furthermore,
define $\varrho' = U\varrho U^{\dagger}$ where $U$ is an arbitrary unitary transformation.
Then
\begin{eqnarray}
r'_i = \Tr \left[ K_i\varrho' \right] = \Tr \left[ U^{\dagger} K_i U \varrho \right] ,
\end{eqnarray}
and thus the transformation of $\varrho$ corresponds to the inverse transformation of
the basis elements $K'_i = U^{\dagger} K_i U$. From the the orthonormality of $\{K_i\}$
we see that $\Tr \left[ K'_iK'_j \right] = \Tr \left[ U^{\dagger} K_i U U^{\dagger}
K_j U \right] = \delta_{ij}$, i.e., $\{K'_i\}$ is also an orthonormal basis.
The transformation of ${\bf r}$ is given by the orthogonal transformation
\begin{eqnarray}
{\bf r}' = R{\bf r},\ R_{ij} = \Tr \left[ K_i K'_j \right] .
\end{eqnarray}
Since $U$ is continuously connected to the identity, $R$ is too, and hence the
transformation is a rotation. The transformation $U$ is also trace-preserving,
which implies that $R$ is restricted to act on a $d^2-1$ dimensional subspace
of $\mathcal{V}$, namely the plane orthogonal to the identity vector ${\bf v}_I$ with
elements $({\bf v}_I)_i = \Tr \left[ K_i \right]$.

A local unitary transformation of a bipartite state
\begin{eqnarray}
\varrho \mapsto \varrho' = U_A\otimes U_B\varrho U_A^{\dagger}\otimes U_B^{\dagger}
\end{eqnarray}
induces the transformation $ M\mapsto M' = R_B M R_A^{\textrm{T}} $, where
\begin{eqnarray}
(R_A)_{ij} =\Tr \left[ U_A ^{\dagger} K^A_i U_A K_j^A \right] , \
(R_B)_{kl} = \Tr \left[ U_B ^{\dagger} K^B_k U_B K_l^B \right] .
\end{eqnarray}
To see that the $\Upsilon_k$'s are invariant under such transformations, it suffices
to note that the above proof of the invariance of the pure state quantities $\Lambda_k$
under unitary transformations, immediately goes through for local rotations of the real
vectors ${\bf a}_i$ and ${\bf b}_i$ representing the hypo- and re-states, respectively.

For any bipartite $\varrho$, there exists a unique Schmidt form
\begin{eqnarray}
\varrho = \sum_i \kappa_i \widetilde{K}_i^A \otimes \widetilde{K}_i^B, \label{eq:mixschmidt}
\end{eqnarray}
where the Hermitian $\{ \widetilde{K}_i^A \}$ and $\{ \widetilde{K}_i^B \}$ are the particular orthonormal
bases of operators on $\mathcal{H}_A$ and $\mathcal{H}_B$ -- corresponding to the pure state Schmidt-bases -- and the real numbers
$\{ \kappa_i \}$ are singular values of $M_{kl} = \Tr \left( K_k^A \otimes K_l^B \varrho \right)$.
In analogy with the pure state case, the invariants $\Upsilon_k^2$ can be seen to be equivalent
to the symmetric polynomials in $\kappa_i^2$, a form of correlation measures similar to those
proposed in Ref. \cite{aniello09}.

Since classical communication can increase correlations, it follows that
$\Upsilon_k$ may increase under LOCC. However, as the following theorem
shows, $\Upsilon_k$ are non-increasing under local operations.

{\it Theorem.} Suppose $\Upsilon_k \mapsto \widetilde{\Upsilon}_k$ under a local
operation
\begin{eqnarray}
\varrho \mapsto \widetilde{\varrho} = \mathcal{E}_{\mathrm{LO}}(\varrho) =
\sum_{ij}A_i\otimes B_j\varrho A_i^{\dagger}\otimes B_j^{\dagger} .
\end{eqnarray}
Then $\widetilde{\Upsilon}_k \leq \Upsilon_k$.

{\it Proof.} We first note that a local operation takes the form
\begin{eqnarray}
M \mapsto \widetilde{M} = S_A M S_B^{\mathrm{T}} ,
\end{eqnarray}
and that the $S$ matrices have a polar decomposition $S = R|S|$, where $R$ is a
rotation and $\left|S\right| = \sum_iq_i {\bf f}_i {\bf f}_i^{\mathrm{T}}, \ 0
\leq q_i \leq 1, \ \mathbf{f}_i^{\mathrm{T}} \cdot \mathbf{f}_j=\delta_{ij}$, is a positive
matrix. Let us first consider the case where $\mathcal{E}_{\mathrm{LO}} = \mathcal{E}_B$
corresponding to $M\mapsto \widetilde{M} = MS_B^{\mathrm{T}} = M |S_B| R_B^{\mathrm{T}}$.
Since we have already proved that $\Upsilon_k$ are invariant under local
unitaries, we may absorb $R_B$ into the choice of hypo-states ${\bf b}_i =
\sum_j b_j^{(i)} \mathbf{f}_j$. Thus, the action of $S_B$ becomes
\begin{equation}
{\bf b}_i \mapsto \widetilde{{\bf b}}_{i} = S_B {\bf b}_i =
\sum_j q_j b_j^{(i)} {\bf f}_j .
\label{eq:S_Bb}
\end{equation}
We further note that $\left| \mathbf{b}_{i_1} \wedge \cdots \wedge {\bf b}_{i_k} \right|^2$
is the norm of the vector $\mathbf{b}_{i_1 \ldots i_k} = \mathbf{b}_{i_1} \wedge \cdots \wedge
{\bf b}_{i_k}$ in the exterior space $ \Omega^k(\mathcal{B})$ of $\mathcal{B}$. The set
$\left\{ {\bf f}_{\mu_1} \wedge \cdots \wedge {\bf f}_{\mu_k} \right\}_{1 \leq \mu_1 < \ldots
< \mu_k \leq d}$ is an orthonormal ordered basis of $\Omega^k (\mathcal{B})$. Thus,
\begin{eqnarray}
\mathbf{b}_{i_1} \wedge \cdots \wedge {\bf b}_{i_k}  = \sum_{1 \leq \mu_1 < \ldots <\mu_k\leq d}
b_{\mu_1 \ldots \mu_k}^{(i_1 \ldots i_k)}
{\bf f}_{\mu_1} \wedge \cdots \wedge {\bf f}_{\mu_k} ,
\nonumber \\
\label{eq:extexpan}
\end{eqnarray}
where
\begin{eqnarray}
b^{(i_1 \ldots i_k)}_{\mu_1\ldots \mu_k} =
\sum_{m_1 \ldots m_k} \epsilon^{\mu_1 \ldots \mu_k}_{m_1 \ldots m_k}
b_{m_1}^{(i_1)} \cdots b_{m_k}^{(i_k)}
\end{eqnarray}
and we may write
\begin{eqnarray}
\upsilon_{i_1 \ldots  i_k}^2 =
\left| \mathbf{b}_{i_1} \wedge \cdots \wedge {\bf b}_{i_k} \right|^2 =
\sum_{1 \leq \mu_1 < \ldots <\mu_k \leq d}
\left| b_{\mu_1 \ldots \mu_k}^{(i_1 \ldots i_k)} \right|^2 .
\label{eq:normext}
\end{eqnarray}
Now, under the local operation $\mathcal{E}_B$, the correlation measure transforms
as $\upsilon^2_{i_1 \ldots i_k} \mapsto \widetilde{\upsilon}^2_{i_1 \ldots i_k} =
\left|\widetilde{\mathbf{b}}_{i_1} \wedge \cdots \wedge \widetilde{\mathbf{b}}_{i_k}\right|^2$,
which can be written as
\begin{eqnarray}
\widetilde{\upsilon}^2_{i_1 \ldots i_k} & = & \left|\left(\sum_{j_1} q_{j_1}b^{(i_1)}_{j_1}
\mathbf{f}_{j_1}\right)\wedge\cdots\wedge\left(\sum_{j_k} q_{j_k}b^{(i_k)}_{j_k}
\mathbf{f}_{j_k}\right)\right|^2
\nonumber \\
& = & \sum_{1\leq\mu_1< \ldots <\mu_k\leq d}q^2_{\mu_1}
\cdots q^2_{\mu_k} \left|b_{\mu_1 \ldots \mu_k}^{(i_1 \ldots i_k)}\right|^2.
\label{eq:long}
\end{eqnarray}
Here, we have used that $q_{\mu}$ are independent of the indices $i_1, \ldots ,i_k$.
Since $0\leq q_{\mu} \leq 1$, it follows that $\upsilon_{i_1 \ldots i_k}$
is non-increasing. Thus, $\widetilde{\Upsilon}_k \leq \Upsilon_k$.

Finally, we need to consider the bi-local operation $M' =
S_A M S_B^{\mathrm{T}}$. This can be written as $M \mapsto M' =
R_A \left| S_A \right|M |S_B| R_B^{\mathrm{T}}$ and from the consecutive application
of the above argument it is clear that
\begin{eqnarray}
\Upsilon_k \geq \widetilde{\Upsilon}_k \geq
\Upsilon_k' ,
\end{eqnarray}
which completes the proof.\begin{flushright}$\Box$\end{flushright}

If we calculate the invariants $\left\{ \Upsilon_k \right\}_{k=2}^{d^2}$ for a pure
state, we expect them to contain redundant information, since the entanglement in
the pure state is characterized by the set $\left\{ \Lambda_k \right\}_{k=2}^{d}$
of pure state invariants. To see how this manifests, consider a pure state with
the Schmidt form $\ket{\psi} = \sum_i\sqrt{p_i}\ket{ii}$. We make particular choice of the local
basis operators, defining them in terms of the local Schmidt bases as
\begin{eqnarray}
E_k & = & \ket{k} \bra{k} , \
k = 1, \ldots ,d,
\nonumber \\
F_{ll'} & = & \frac{1}{\sqrt{2}}\left(\ket{l} \bra{l'} + \ket{l'} \bra{l} \right) , \
1 \leq l < l'
\leq d,
\nonumber \\
G_{mm'} & = & \frac{i}{\sqrt{2}}\left(\ket{m} \bra{m'} - \ket{m'} \bra{m} \right) , \
1 \leq m < m' \leq d.\label{eq:op_basis}
\end{eqnarray}
The Hilbert-Schmidt representation $M_\psi$ of the state $\ket{\psi}\bra{\psi}$ is
diagonal in this basis, with the diagonal values given by
\begin{eqnarray}
e_k = p_k & = & \Tr \left[ E_k^A\otimes E_k^B\ket{\psi}\bra{\psi} \right] ,
\nonumber \\
f_{ll'} = \sqrt{p_lp_{l'}} & = & \Tr \left[ F_{ll'}^A\otimes F_{ll'}^B\ket{\psi}\bra{\psi} \right],
\nonumber \\
g_{mm'} = -\sqrt{p_mp_{m'}} & = & \Tr \left[ G_{mm'}^A\otimes G_{mm'}^B\ket{\psi}\bra{\psi} \right].\label{eq:mixschmidtcoef}
\end{eqnarray}
Note that this is essentially the mixed state Schmidt decomposition given in Eq. (\ref{eq:mixschmidt}) with mixed state Schmidt coefficients $\left\{ e_k,f_{ll'},g_{mm'} \right\}$, i.e., we have that
\begin{eqnarray}
\ket{\psi}\bra{\psi} = \sum_{k=1}^d p_k E_k^A\otimes E_k^B + \sum_{1\leq l<l'\leq d} \sqrt{p_lp_{l'}}\left( \ F_{ll'}^A\otimes F_{ll'}^B - \ G_{ll'}^A\otimes G_{ll'}^B\right).
\end{eqnarray}

The measures $\Upsilon_k$ are functions of the mixed state Schmidt coefficients  $\left\{ e_k,f_{ll'},g_{mm'} \right\}$, which for pure states are, in turn, simple functions of the pure state Schmidt coefficients, as can be seen from Eq. (\ref{eq:mixschmidtcoef}). Hence $\Upsilon_k$ can be
expressed in terms of the pure state measures $\Lambda_k$. For some $k$ this relation becomes
simple, e.g., one can show that
\begin{eqnarray}
\Upsilon_2^2 = 2\left( \Lambda_2^2 - \Lambda_2^4 \right), \ \
\Upsilon_3^2 = 2\left( \Lambda_2^4 - \Lambda_2^6 \right), \ \
\Upsilon^2_{d^2} = \Lambda_d^{2d}.
\end{eqnarray}

\section{Application: Quantum dynamics}
\label{sec:application}
We illustrate the correlation measures $\Upsilon_k$ by looking at how the correlations
of a maximally entangled state $\ket{\psi} = \frac{1}{\sqrt{d}} \sum_{i=1}^d\ket{ii}$
of a $d \times d$ dimensional system changes under two types of decoherence.

We first consider the depolarization channel $\mathcal{E}$ defined as the map
\begin{eqnarray}
\psi \mapsto \varrho_W = \mathcal{E} (\psi) =  p\psi + (1-p)\varrho_{\ast}
\label{eq:depol}
\end{eqnarray}
of $\psi$. Here, $\psi = \ket{\psi}\bra{\psi}$ and $\varrho_{\ast} = \frac{1}{d^2}\Id_{AB}$,
i.e., the output $\varrho$ is a Werner state that connects the maximally entangled
state $\psi$ for $p=1$ and the random mixture $\varrho_{\ast}$ for $p=0$. The relative
state map $\mathfrak{L}_W$ induced by $\varrho$ acts on a hypo-state $\tau\in\mathcal{S}_A$ as
\begin{eqnarray}
\tau \mapsto \pi = \mathfrak{L}_W (\tau) = p\mathfrak{L}_\psi(\tau) + (1-p)\mathfrak{L}_{\ast} (\tau),
\end{eqnarray}
where $\mathfrak{L}_\psi$ and $\mathfrak{L}_{\ast}$ are the maps induced by $\psi$ and
$\varrho_{\ast}$, respectively. In particular, for any $Y \in \mathcal{B}(\mathcal{H}_A)$,
we find $\mathfrak{L}_{\ast} (Y) = \frac{1}{d^2} \Tr \left[ Y \right] \Id_B$, i.e., the
relative state map defined by $\varrho_{\ast}$ maps any element of $\mathcal{B}(\mathcal{H}_A)$
to an operator proportional to the reduced state of subsystem $B$.

\begin{figure} [htbp]
\centerline{\epsfig{file=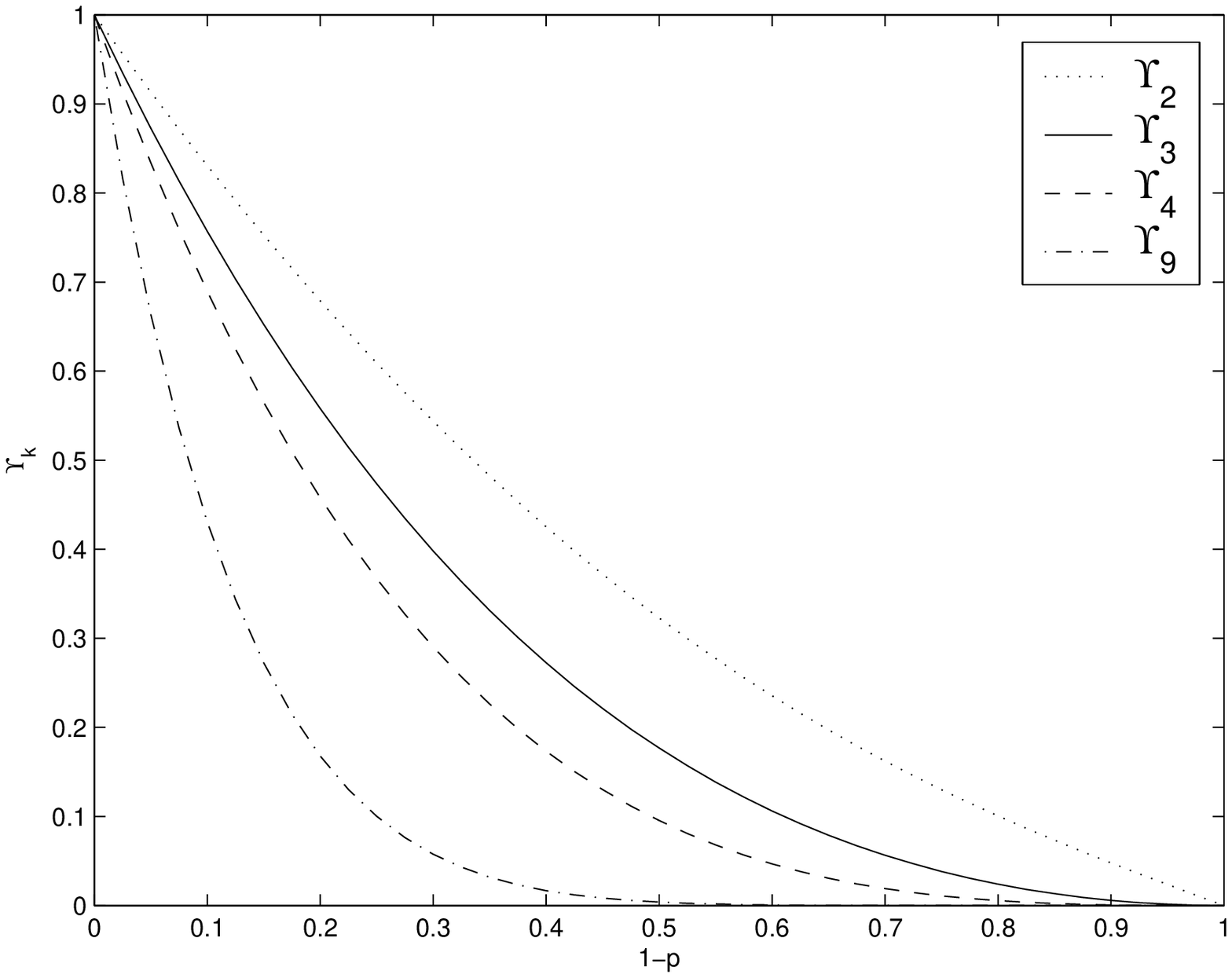, width=8.5 cm}}
\vspace*{13pt}
\fcaption{\label{fig:depolarization}The invariants of a maximally entangled $3\times 3$
state undergoing depolarization. With increasing noise, all $\Upsilon_k$ tend to zero.}
\end{figure}

Now, let $\{K_i^A\}$ and $\{K_i^B\}$ be orthonormal bases of $\mathcal{B}(\mathcal{H}_A)$ and
$\mathcal{B}(\mathcal{H}_B)$, respectively, with the additional property that $K_1^A =
\frac{1}{\sqrt{d}} \hat{1}_A$ and $K_1^B = \frac{1}{\sqrt{d}} \hat{1}_B$. This implies
$\Tr \left[ K_i^A \right] = \Tr \left[ K_i^B \right] = 0$ for $ i>1$. By choosing
$\tau_i = K_i^A$, we obtain the re-states
\begin{eqnarray}
\pi_1 & = & \mathfrak{L}_W (\tau_1) = \left(\frac{1}{d}\right)^{\frac{3}{2}}\Id_B ,
\nonumber \\
\pi_i & = & \mathfrak{L}_W (\tau_i) = p\mathfrak{L}_\psi(\tau_i)  = p\pi'_i , \ i>1.
\end{eqnarray}
Let us use these expressions to evaluate first $\Upsilon^2_2$ explicitly. The individual
terms are given by $\upsilon_{i_1i_2} = |\pi_{i_1} \wedge \pi_{i_2}|$, which can take
two values
\begin{eqnarray}
\upsilon_{1i_2} & = &
|\pi_1 \wedge \pi_{i_2}| =  \frac{p}{d^2}, \ 1<i_2,
\nonumber \\
\upsilon_{i_1i_2}& = &
|\pi_{i_1}\wedge \pi_{i_2}| = \frac{p^2}{d^2}, \ 1<i_1<i_2,
\end{eqnarray}
where we have used that $|\pi'_i \wedge \pi'_j| = 1/d^2$ for all $i\neq j$. Hence, we have
that
\begin{eqnarray}
\Upsilon_2^2 = \binom{d^2}{k}^{-1}p^2\left(d^2-1 + \binom{d^2-1}{2}p^2\right) =
\frac{p^{2}}{d^2}\left(2 + (d^2-2)p^2\right).
\end{eqnarray}
Generalizing to arbitrary $k=2, \ldots ,d^2$ yields
\begin{eqnarray}
\Upsilon_k^2  =
d^{2k}\binom{d^2}{k}^{-1}\sum_{1\leq i_1<...<i_k\leq d^2} \upsilon^2_{i_1...i_k} =
p^{2(k-1)}\left[ \frac{k}{d^2} + \left( 1-\frac{k}{d^2} \right) p^2 \right]
\end{eqnarray}
which vanishes when $p \rightarrow 0$. The $k=2,3,4,9$ invariants are shown
in Fig. \ref{fig:depolarization} for $d=3$.

\begin{figure} [htbp]
\centerline{\epsfig{file=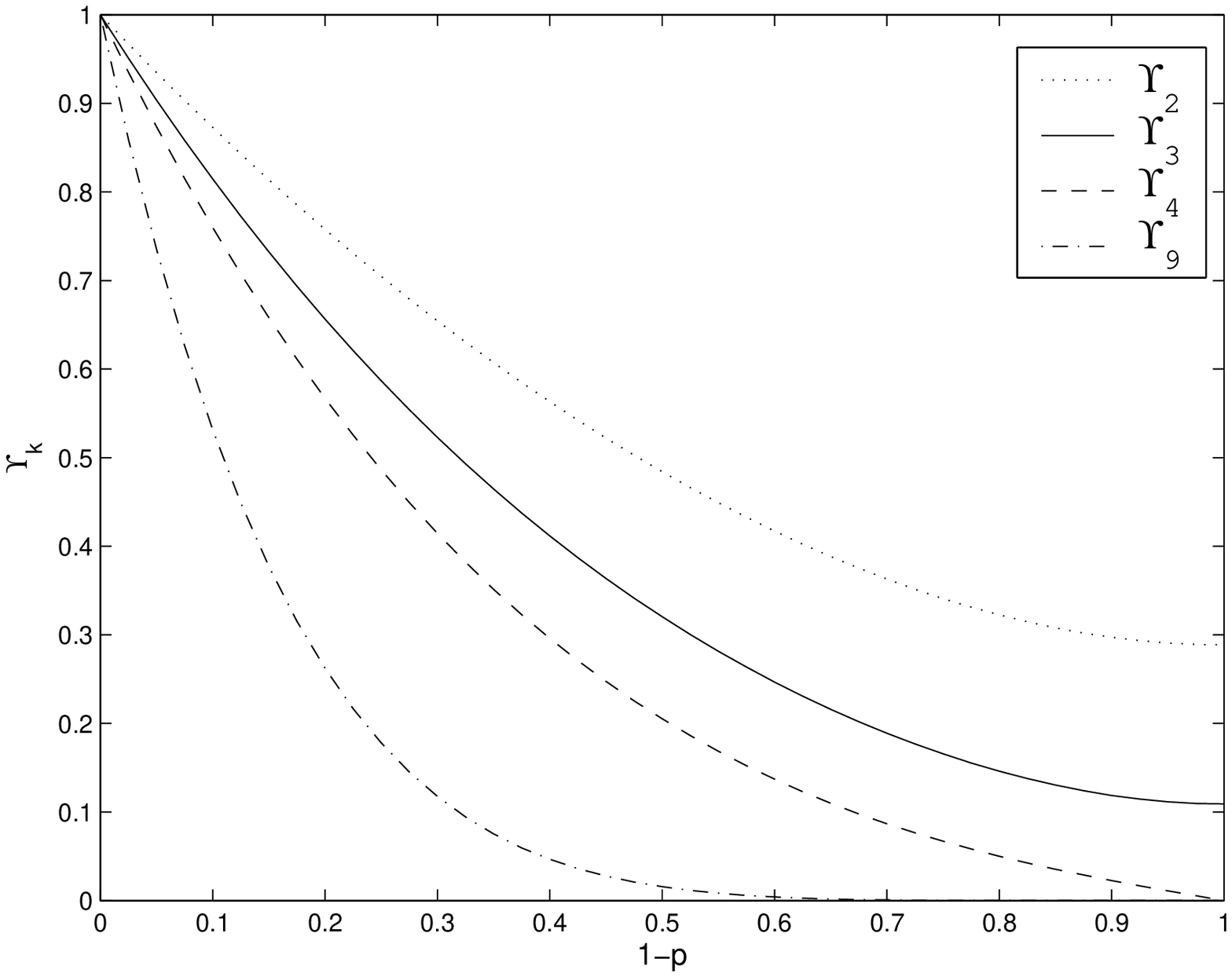, width=8.5 cm}}
\vspace*{13pt}
\fcaption{\label{fig:dephasing}The product basis decoherence of a maximally entangled state.
Note that $\Upsilon_k \rightarrow 0$ when $p \rightarrow 0$ for $k>d$, while it goes to a
finite value for $k\leq d$. The final state defined by $\mathcal{F}$ for $p=0$ is a maximally
correlated separable state.}
\end{figure}

Secondly, we look at how the invariants change under product-basis decoherence of the maximally
entangled state $\psi$. Let the product basis be composed of the local Schmidt-bases, i.e.,
$E_i^A \otimes E_j^B = \ket{i} \bra{i} \otimes \ket{j} \bra{j}$. The channel $\mathcal{F}$
can be represented as
\begin{eqnarray}
\psi \mapsto \varrho_D = \mathcal{F} (\psi) =
p\psi+ (1 - p)\sum_{ij=1}^d E_i^A\otimes E_j^B\psi E_i^A\otimes E_j^B.
\end{eqnarray}
If we define the maximally decohered state $\Xi = \sum_{ij} E_i^A \otimes E_j^B \psi
E_i^A \otimes E_j^B = \frac{1}{d} \sum_i E_i^A \otimes E_i^B$, then
\begin{eqnarray}
\mathfrak{L}_D (\tau_i) = p\mathfrak{L}_\psi(\tau_i) + (1-p)\mathfrak{L}_\Xi (\tau_i)
\end{eqnarray}
We choose the local basis operators given in Eq.(\ref{eq:op_basis}). With the identification
$\{\tau_i\} = \{E_k,F_{ll'},G_{mm'}\}$, we have $\mathfrak{L}_\psi (\tau_i) =
\mathfrak{L}_\Xi(\tau_i)$ for $1\leq i \leq d$, and $\mathfrak{L}_\Xi(\tau_i) = 0$
for $i> d$. $\mathfrak{L}_D (\tau_i)$ takes two values in terms of $\mathfrak{L}_\psi (\tau_i)$:
\begin{eqnarray}
\mathfrak{L}_D (\tau_i) = \bigg\{
\begin{array}{ll}
\mathfrak{L}_\psi (\tau_i), & 1 \leq i \leq d \\
p\mathfrak{L}_\psi (\tau_i), & d < i \leq d^2 \\
\end{array}.
\end{eqnarray}
It is then a matter of combinatorics to show that the invariants for general $k$ are
given by
\begin{eqnarray}
\Upsilon_k^2 =  \binom{d^2}{k}^{-1} \sum_{l=0}^k \binom{d}{k-l} \binom{d^2-d}{l}p^{2l} .
\end{eqnarray}
The $k=2,3,4,9$ invariants are shown in Fig. \ref{fig:dephasing} for $d=3$.

For the depolarization channel, all $\Upsilon_k \rightarrow 0$ when $p \rightarrow 0$, i.e.,
all $k$-level correlations are suppressed and vanish for the final product state. However, the
Werner state for $p\neq 0$ inherits the symmetry of the maximally entangled state: for any choice
of orthonormal measurement basis $\{\ket{a_i}\}$ (i.e., an observable) at site $A$, there is a
corresponding orthonormal basis $\{\ket{b_i}\}$ at site $B$, in which the correlations (as measured
by, e.g., mutual information) exhibited by the resulting probability distribution will be non-zero.
In terms of the invariants, this is due to that $\Upsilon_{d^2} \neq 0$, or, in terms of the
$k$-vectors, that for any choice $\{\ket{a_i}\}$ the objects $\mathfrak{L}_W (\ket{a_1}\bra{a_1})
\wedge \cdots \wedge \mathfrak{L}_W(\ket{a_d}\bra{a_d})$ and $\mathfrak{L}_W (\tau_1) \wedge \cdots
\wedge \mathfrak{L}_W(\tau_{d^2})$ have a $d$-dimensional intersection.
\begin{figure} [htbp]
\centerline{\epsfig{file=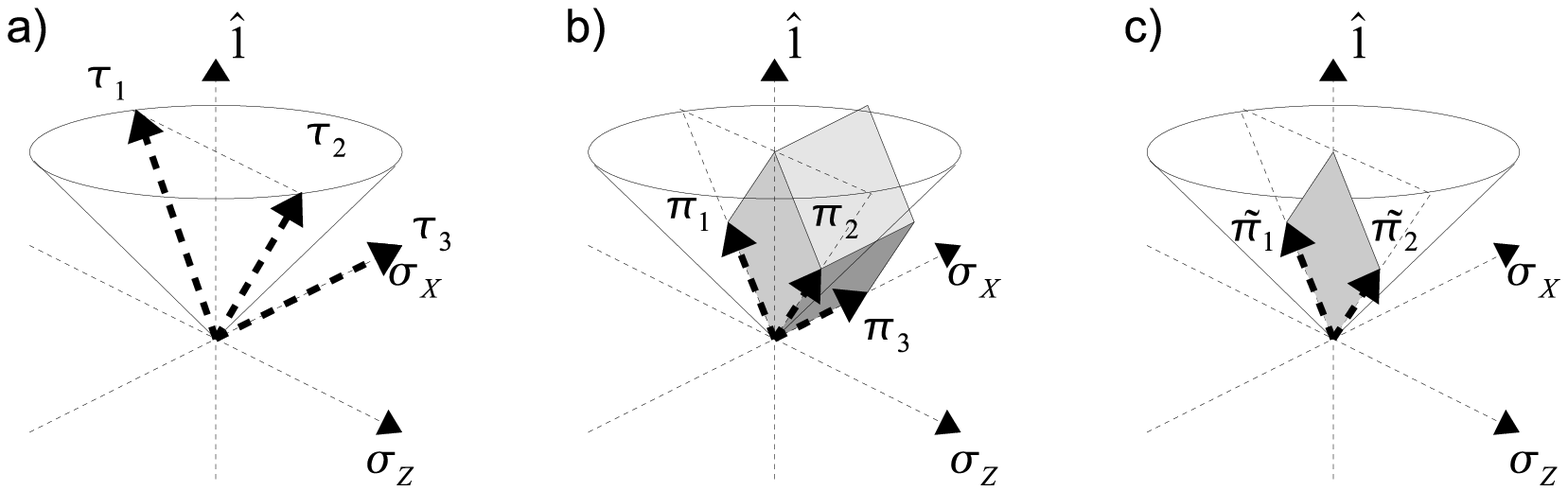, width=12 cm}}
\vspace*{13pt}
\fcaption{\label{fig:mixed}An illustration of the mixed state relative states and the mixed state
correlation measures for the $2\otimes 2$ system. In a) we show a three dimensional subspace of
$\mathcal{B}(\mathcal{H}_A)$, spanned by basis operators $K^A_1 = \Id /\sqrt{2}$, $K^A_2 =
\sigma_x/\sqrt{2}$ and $K^A_3 = \sigma_z/\sqrt{2}$, and b) and c) shows the corresponding
subspace of $\mathcal{B}(\mathcal{H}_B)$. The disc enclosed by the circle orthogonal to $\Id$
is a subspace of the state space $\mathcal{S}$, i.e., the $xz$-plane of the Bloch-sphere, with
the pure states on the boundary. The corresponding subspace of subnormalized states $\mathcal{S}'$
is the cone with the circle as its base.  Three (out of four) hypo-states $\tau_1 =
\ket{0}\bra{0},\ \tau_2 = \ket{1}\bra{1},\ \tau_3 = K_2$ are shown in a). The re-states
$\pi_i = \mathfrak{L}_\psi(\tau_i)$, defined by the maximally entangled state $\psi =
\frac{1}{2}\sum_{k,l=0}^1\ket{kk}\bra{ll}$, is shown in b), where the volume spanned is
$\upsilon_{123} = |\pi_1\wedge\pi_2\wedge\pi_3| = (1/2)^3$. In c) the restates
$\widetilde{\pi}_i=\mathfrak{L}_\Xi(\tau_i)$ of the maximally classically correlated state
$\Xi = \frac{1}{2}(\ket{00}\bra{00} + \ket{11}\bra{11})$, is shown, where $\upsilon_{123} =
|\widetilde{\pi}_1 \wedge \widetilde{\pi}_2 \wedge \widetilde{\pi}_3 | = 0$ since
$\widetilde{\pi}_3 = 0$. The quantity $\upsilon_{12} = 1/4$ is the only nonzero contribution
to the invariants, and hence $\Upsilon_4 = \Upsilon_3 = 0$ and only $\Upsilon_2 \neq 0$,
which characterizes a two-qubit 0-discord state.}
\end{figure}

For the product-basis decoherence channel, which is of interest as, e.g., a model for measurement
einselection,  we see that $\Upsilon_k \rightarrow 0$ for $k>d$ and $\Upsilon_k \rightarrow
\binom{d^2}{k}^{-1/2}\binom{d}{k}^{1/2}$ for $k\leq d$,  when $p \rightarrow 0$ (see Fig.
\ref{fig:dephasing}). This can be related to proposed measures of classical correlations
in quantum states \cite{hendersonvedral,ollivier2001}, in particular, quantum discord
\cite{ollivier2001} defined as
\begin{eqnarray}
D(A:B|\left\{ \tau_i \right\}) = S(\rho_B)- S(\rho) +\sum_ip_iS(\pi_i/p_i),
\end{eqnarray}
where $p_i = \Tr\left[ \pi_i \right]$, $S(\rho)$ denotes the von Neumann entropy of the state
$\rho$, and we have the restriction $\tau_i\in\mathcal{S}(\mathcal{H}_A)$, which ensures that
$\mathfrak{L}_\varrho(\tau_i)=\pi_i\in\mathcal{S}'(\mathcal{H}_B)$ and that
$\pi_i/p_i\in\mathcal{S}(\mathcal{H}_B)$. This definition also utilizes a relative state
construction in that they are derived from entropies over subsystem $B$ that are conditioned
on measurements on subsystem $A$. The minimum discord of a state
\begin{eqnarray}
D_{\min}(A:B) = D(A:B|\left\{ \tau_i \right\}_{\min}) = S(\rho_B)- S(\rho) +
\min_{\left\{ \tau_i \right\}}\left(\sum_ip_iS(\pi_i/p_i)\right)
\end{eqnarray}
quantifies the amount of information lost in the optimal correlation measurement.

The final state $\Xi$ is the maximally correlated separable state and likewise the maximally
correlated zero-discord state. A zero-discord state is characterized by that $D_{\min} = 0$,
which means that all its correlations can be extracted by a single measurement setup, and
consequently that the state is robust under this particular measurement, and that the classical
mutual information over the probability distribution obtained by measuring in a product basis
equals the quantum mutual information of the state. This can be related to the invariants in
the following way: If $\Upsilon_k \neq 0$ for $k>d$, then $D_{\min}(A:B)\neq 0$, and if
$D_{\min}(A:B)=0$, then $\Upsilon_k = 0$ for $k>d$. The first implication we understand as
that the re-states span a $k>d$-dimensional subspace of $\mathcal{S}_B$, while the (complete)
set of projectors constituting a measurement basis only span a $d$-dimensional subspace, thus
the $k$-volume spanned by the re-states ''collapses'' into a $d$-volume upon measurement,
which is what we see in the example of product basis decoherence. The restates of the pre-
and post-measurement states $\psi$ and $\Xi$ are shown in Figs. \ref{fig:mixed}b and
\ref{fig:mixed}c, respectively. Conversely, the second implication illustrates that
the re-states of a zero-discord state, which is robust under some product basis measurement,
can maximally span a $d$-dimensional subspace. It also follows that, contrary to the
symmetric Werner state, that one can find a set of $d$-hypo-states $\left\{  \ket{a'_i} \right\}$
(bases $\left\{  \ket{a'_i} \right\}$ and $\left\{  E_i \right\}$ are mutually unbiased) such
that $\mathfrak{L}_\Xi(\ket{a'_i}\bra{a'_i}) = \mathfrak{L}_\Xi(\ket{a'_j}\bra{a'_j})$,  and
none of the state's correlations can be extracted.

\section{Conclusions}
The concept of relative state, originally developed by Everett \cite{everett57} to deal
with the measurement problem in quantum mechanics, has been used to construct measures
of correlations in pure bipartite quantum states of arbitrary dimension. The basic idea
is to quantify how much information one observer can obtain about measurements that
can be performed by another observer, if they are allowed only to do local, projective
measurements. These correlation measures have been shown in detail to be invariant under
local unitary transformations of the shared bipartite state. We have further shown that
the present correlation measures coincide with those given by concurrence hierarchies
\cite{fan03} and $I$ concurrence \cite{rungta01}, providing an alternative operational
interpretation of these measures.

We have extended the notion of relative state to generalized measurements. This allows
for studies of the correlation structure of mixed bipartite states. The corresponding
measures quantify the total correlation in the sense that they vanish for product states,
and are non-increasing under local operations, but may increase under LOCC. We have
illustrated the behavior of the mixed state correlation measures for bipartite systems
of arbitrary dimension undergoing two different types of open system dynamics.

\section*{Acknowledgments}
PR acknowledges financial support from the G\"oran Gustafsson Foundation. ES acknowledges
support from the National Research Foundation and the Ministry of Education (Singapore).

\end{document}